\documentclass[prb,aps,floats,amssymb,showkeys,showpacs,
superscriptaddress,floatfix]{revtex4}
\usepackage{graphicx}
\usepackage{graphics}
\usepackage{epsfig,bm}
\usepackage{rotating}
\usepackage{float} 
\begin{document}

\title{
Ising low-temperature polynomials  and hard-sphere  
 gases on cubic lattices of general dimension }

\author{P. Butera}
\email{paolo.butera@mib.infn.it}
\affiliation
{Dipartimento di Fisica Universita' di Milano-Bicocca\\
and\\
Istituto Nazionale di Fisica Nucleare \\
Sezione di Milano-Bicocca\\
 3 Piazza della Scienza, 20126 Milano, Italy}
\author{M. Pernici}
\email{mario.pernici@mi.infn.it} 
\affiliation
{Istituto Nazionale di Fisica Nucleare \\
Sezione di Milano\\
 16 Via Celoria, 20133 Milano, Italy}

\date{\today}
\begin{abstract}
  We derive and analyze the low-activity and low-density expansions of
  the pressure for the model of a hard-sphere gas on cubic lattices of
  general dimension $d$, through the 13th order.  These calculations
  are based on our recent extension to dimension $d$ of the
  low-temperature expansions for the specific free-energy of the
  spin-1/2 Ising models subject to a uniform magnetic field on the
  (hyper-)simple-cubic lattices.  Estimates of the model parameters
  are given also for some other lattices.
\end{abstract}

\maketitle

\section{Introduction}
A simpler and mathematically more tractable discretization 
 of the long-studied hard-sphere
fluid in continuum space  is called ``hard-sphere lattice gas'' 
(HSLG) with nearest-neighbor exclusion\cite{YL1,YL2,burley,fisgaunt,gaunt}.
 In this model the  sites of a regular lattice are the only
allowed positions of the constituents. 
The pair interaction potential is $+\infty$ for constituents centered
on nearest-neighbor sites and vanishes otherwise.  Double occupancy of
sites is forbidden. In particular in $2d$, on a square ($sq$) lattice,
the spheres of the model can be viewed as ``hard-squares'' oriented at
$\pi/4$ with respect to the lattice axes and whose diagonals equal two
lattice spacings. In $3d$ the spheres are effectively ``hard octaedra'' 
and analogously for higher-dimensional lattices, they are polytopes (i.e.
convex hyper-polyhedra) whose vertices are the nearest neighbors of
the sites on which they are centered.

If the properties of the HSLG  on a
finite lattice are described by the statistical mechanics formalism of
the grand partition function, it turns out that the $n$th coefficient of
 the low-activity(LA)
series-expansion  of the grand-potential   has the
same combinatorial definition\cite{burley,fisgaunt,gaunt}
 as the coefficient of the highest
power of the temperature-like variable $u$ in the $n$th low-temperature
(LT) polynomial of a ferromagnetic Ising model with spin $S=1/2$ on the
same lattice.  This relationship remains valid when the
thermodynamic limit is taken.  When this fact was
recognized, it became trivial to write
down the LA expansions of the pressure once the LT expansion of the 
Ising model on the same lattice, in presence of a magnetic field, is known.

Thus, for the one-dimensional lattice, one can write a LA expansion
valid to all orders.  For the $sq$ lattice, initially the LA expansion
could be obtained\cite{burley,fisgaunt,gaunt} in this way only through
the 12th order. The derivation was later extended\cite{sykes2d3d}
through the 21st order.  Finally, by a more powerful approach based on
the corner-transfer-matrix method\cite{baxter}, the LA series was
pushed through the 42nd order and more recently\cite{chan} through the
92nd order.

For the triangular lattice (``{\it hard-hexagon}'' model), an exact
solution\cite{baxtri,baxbook} of the HSLG model was devised. It is
expected\cite{bax} that also the honeycomb ($hc$) lattice case 
(called the ``{\it hard-triangle}'' model) is soluble, however,
presently we know only a a 25th order LA expansion derived from the LT
Ising expansion\cite{syk2d,syk25hc} for the $hc$ lattice.  For
lattices of dimension $d>2$, the transfer-matrix techniques are not
efficient, so that one has to rely only on the LT Ising expansions and
no LA data of extension comparable with that of the $sq$ lattice can
be derived.  In the case of the three-dimensional hydrogen-peroxide
($hpo$) lattice, a bipartite cubic lattice of coordination number 3,
the known Ising LT series\cite{sykhpo} yields a 23rd order LA series
for the HSLG model.  For the three-dimensional bipartite $diamond$
lattice, with coordination number 4, the first 17 LA coefficients can
be obtained\cite{syk25}. Only the first 15 Ising LT polynomials are
presently known\cite{sykes2d3d,sykes4d} for the simple-cubic ($sc$)
lattice in $3d$ and the hyper-simple-cubic ($hsc$) in $4d$, while the
first 11 polynomials\cite{sykesbcc} are known in the case of the
body-centered-cubic ($bcc$) lattice. No LT data at all existed up to
now for the $hsc$ and the hyper-body-centered-cubic ($hbcc$) lattices
of dimension $d>4$.

We have recently calculated\cite{BP}  the Ising LT
polynomials  for all $hsc$ 
lattices in general dimension $d$ through the 13th order. 
  As a result, in
this paper we can present the LA and low-density (LD)
 expansions for the HSLG through the same order,
  having observed
  that the LD expansion coefficients $v_k(d)$ of the pressure  are
  polynomials of degree $[\frac{k}{2}]$ in $d$.  It is worth noticing
  that also the LD expansion coefficients of the pressure for the
  dimer model on the $hsc$ lattices share a similar (slow) polynomial
  dependence on $d$, making it  possible to determine\cite{bfp}
  the LD expansions through the 20th order.

  We shall analyze the LA expansions for the HSLG to extract
  information about the properties of the model for $d>4$, which, up
  to now, have been the subject of a single accurate MonteCarlo(MC)
  study\cite{hering}. To support some of the indications suggested
  from our study for the higher-dimensional lattices, we shall also
  perform an analysis of the LA and the LD series for the $hc$, the
  $hpo$ and the $diamond$ lattices, to which little attention was devoted
  in the literature.

It is also well known that, using the standard
correspondence\cite{YL2} between the lattice gas and the spin-1/2
Ising model, one can relate the HSLG with an LT anti-ferromagnetic
Ising spin system subject to a magnetic
field\cite{temp,burley,metcalf} on the same lattice and thus further
insight can come also from this standpoint.  In particular, in analogy
with the LA series, the high-activity (HA) expansions coefficients (in
powers of $1/z$) for the pressure can be directly read\cite{fisgaunt}
from the Ising LT anti-ferromagnetic polynomials. Unfortunately, the
LT anti-ferromagnetic data are scarce and therefore direct
methods\cite{fisgaunt,gaunt} have been necessary to compute the HA
series\cite{baxter}, presently known only through order 23 for the
$sq$ lattice, and through the 16th or the 19th orders, respectively for
the simple-cubic ($sc$) and the body-centered-cubic ($bcc$) lattices.
No such series exist for the lattices with $d>4$, with which we shall
be mainly concerned in this paper.

  For the HSLG model  one is mainly interested into

  {\it a)} the leading {\it nonphysical} singularity\cite{groene} of
  the pressure located at a small negative value of the activity. Its
  exponent is known to be simply related to that of the universal
  \cite{YL1,fishuni,laifish} Yang-Lee edge-singularity for spin systems of
  the same dimension, as well as to the exponents of several other
  systems\cite{bperyl}, such as the directed branched polymers, the
  undirected site or bond animals etc..

  {\it b)} the parameters of the expected {\it physical}
  phase-transition which takes place as the density increases and
  changes the LD disordered phase into an HD ordered one. In this
  transition, associated with the nearest singularity of the pressure
  on the positive activity (or density) axis and accurately
  checked\cite{hering,bloetez,bloeteuniv} to be Ising-like for the
  bipartite lattices (to which we shall restrict this study), one of the
  two equivalent sublattices becomes preferentially occupied at HD,
  while at LD (or equivalently at LA), the constituents are uniformly
  distributed all over the lattice sites. We may also remark\cite{frenkel}
  that the transition is ``entropy-driven'', because the internal
  energy vanishes for the allowed configurations of the system and the
  temperature turns out to be an irrelevant constant, so that the free
  energy coincides up to a sign with the entropy.

  The paper is organized as follows. In Section II, we recall the
  structure of the LT expansion for a ferromagnetic spin 1/2 Ising
  model and write down the corresponding LA and LD expansions of the
  pressure for the HSLG.  The analysis of these expansions for
  lattices of various dimensions, leads to a conjecture on the nature
  of the nearest singularity in the complex density plane for $d \ge
  3$, and is presented in Section III. Simple estimates of the entropy
  constants for lattices of not too high $d$ are discussed in Section
  IV.  The last Section contains a summary of our results.

  The Appendix A0 reviews in some detail our derivation of the LT
  polynomials of the spin 1/2 Ising model subject to a magnetic field,
  from the corresponding HT  expansion.

  In the Appendix A1, we argue  that the virial expansion coefficients
$v_k(d)$ can be expressed as polynomials in $d$  of degree
  $[\frac{k}{2}]$ and present 
   handy expressions of the LA and the LD expansions
  of the pressure valid for $hsc$ lattices of general dimension. 

\section{ Ising  LT expansions and HSLG LA expansions}
On a finite $d$-dimensional lattice of $N$ sites (we set the lattice
spacing $a=1$), the spin-$1/2$ ferromagnetic Ising model in an
external magnetic field $H$, is described by the
Hamiltonian\cite{wortis,isingesse}
\begin{equation}
{\cal H}_N\{\sigma\}=-J  \sum_{<ij>} \sigma_i \sigma_j-mH\sum_i \sigma_i
\label{isiesse}
\end{equation}
where $\sigma_i=-1,+1$ denotes the spin variable at the site $\vec i$,
$J>0$ is the exchange energy and $m$ (set equal to one in what follows) 
is the magnetic moment of a spin.
The first sum in Eq. (\ref{isiesse}) extends over all distinct
nearest-neighbor pairs of sites, the second sum over all lattice
sites.  If we set $\beta=1/k_BT$, $K=J \beta$, with $k_B$ the
Boltzmann constant and $T$ the temperature, while $h=H \beta$ denotes
the reduced magnetic field, then in the LT  and high-field limit, we
can write the free energy per site $-\beta {\cal F}_{LT} (K,h)=
\lim_{N \to \infty} N^{-1} {\rm ln} Z_N(K,h)$ as
\begin{equation}
-\beta{\cal F}_{LT}(K,h)=h+\frac{1}{2}qK+\sum_{n=1}^{\infty}
 L^{(1/2)}_n(u)\mu^n
\label{fl0}
\end{equation}
 Here $u=exp[-4K]$, $\mu=exp[-2h]$, and $q$ is the coordination number
 of the lattice.  The series-expansion coefficients of the free energy in
 powers of $\mu$, denoted by
 $L^{(1/2)}_n(u)$, 
 are polynomials in the variable $u$.

 We have recently derived (or, in some cases, extended) the LT
 expansions of the specific free-energy for the spin-$S$ Ising models,
 on $hsc$ and $bcc$ lattices of arbitrary dimension $d$ in presence of a
 magnetic field.  Our additional results were not obtained by
 extending the direct graphical calculation, but more simply from our
 high-temperature (HT)  expansions\cite{BP}, by
 performing on these an appropriate transformation to the  LT and
 high-field variables.  In particular, we have derived 13th order LT
 expansions on the $hsc$ lattice of general dimension $d$, for
 $S=1/2$.

 As anticipated in the introduction, the knowledge of these expansions
 yields immediately\cite{fisgaunt,gaunt} the coefficients of the LA
 (LD) expansion for the grand-potential of the HSLG on the same
 lattices.

 For convenience, let us now state a few standard definitions from
 the statistical mechanics description of the HSLG model.  Due to the
 nearest-neighbor exclusion interaction, the $r$-th coefficient
 $g_r(N)$ of the LA expansion of the grand partition-function
 $\Xi_N(z)$  for a $N$-site lattice simply counts the
 allowed configurations of $r$ hard spheres.  In the absence of
 non-hard-core interactions and thus of an energy scale for the
 system, the statistical mechanics is temperature independent,
 i.e. the system is ``athermal'', and one can simply set $k_BT \equiv
 1$. The first three coefficients of the LA expansion of $\Xi_N(z)$
 depend on the lattice structure and size $N$ as simply as
\begin{equation}
\Xi_N(z)= \sum_r g_r(N,d)z^r= 1+Nz+N(N-q-1)z^2/2!+...
\end{equation}
where $z=exp(\mu)$ and $\mu$ is the chemical potential.  It is worth
noting that the coefficients $g_r(N,d)$ enumerate the ``independent
sets'' of $r$ vertices in the $N$-vertex graph induced by the (finite)
lattice under study. Thus, in combinatorial language $\Xi_N(z)$ is the
``independent-set generating polynomial'' of the lattice.  Computing
these quantities for general graphs has been a much studied problem in
combinatorial mathematics\cite{hboo}.

   In the thermodynamic limit, the coefficients
of the LA expansion\cite{mayer} of the ``grand-potential'' (or
``pressure'') per site
\begin{equation}
\lim_{N \to \infty} \frac{1}{N} {\rm ln} \Xi_N(z)  =
 p(z) =\sum_rc_r(d)z^r =z-\frac{1}{2}(q+1)z^2+...
\end{equation} 
 are formally obtained by picking out the coefficients of the terms
linear in $N$ in the expansion of ${\rm ln} \Xi_N(z)$.  In the same
limit, the number density of the gas is defined by
\begin{equation}
\rho(z)=\lim_{N \to \infty} \langle n \rangle/N= z\frac{d p(z)}{dz}=z-(q+1)z^2+..
\label{dens}
\end{equation} 
with $\langle n \rangle$ the mean number of hard spheres on the lattice. On the
bipartite lattices that we shall consider here, the physical range of
the density $\rho$ is $[0,1/2]$.

 The virial expansion of the pressure (LD expansion) is then obtained by using
 Eq. (\ref{dens}) to express the activity $z$ as function of $\rho$ and
 substituting it in $p(z)$
\begin{equation}
p(\rho)=\sum_rv_r(d)\rho^r=\rho+\frac{1}{2}(q+1)\rho^2+...
\end{equation} 
 The  compressibility $K_T$ defined as 
\begin{equation}
 \rho  K_T = \frac {d\rho}{dp}
\end{equation} 
and the specific entropy
\begin{equation}
S=-\rho {\rm ln}z +p
\label{entro}
\end{equation}
are also of interest.  Note that $S$ vanishes not only as $z \to 0$,
but also as $z \to \infty$, so that there is no residual entropy at
maximum density, since for the bipartite lattices, independently of
their structure, the first terms of the HA expansion\cite{fisgaunt}
for the pressure and the density are $ p = \frac{1} {2} {\rm
  ln}z+\frac{1} {2z} +O(\frac{1} {z^2})$ and $\rho(z)= 1/2 -\frac{1}
{2z}+O(\frac{1} {z^2}) $, respectively.

In the case of the $hsc$ lattices, the coefficients $[n,i](d)$ of the
various powers of $u$ in the LT polynomials $L^{(1/2)}_n(u)$ can be
represented as simple polynomials in $d$.  This property of the LT
polynomials reflects the analogous property\cite{isigend} of the
 HT expansion of the Ising model for this class of lattices.
In the Appendix A1, we have discussed why it is so. Correspondingly,
we can write simple exact expressions, valid for any $d$, of each
coefficient in the LA and in the LD expansions of the pressure.

These general $d$ expansions for the $hsc$ lattices, that we regard as
the main result of this paper in spite of their still moderate length, are
reported in the Appendix A1.  No such simplification is possible for
the $hbcc$ lattices.  In this case, distinct expansions must be
written for each value of $d$ and some of them will be tabulated
elsewhere.

In the case of the one-dimensional lattice,  the
free-energy of the Ising model in a field can be computed to all
orders yielding exact expressions for the LA expansion of the HSLG
model:
\begin{equation}
 p(z)={\rm ln}\big(\frac {1+\sqrt{1+4z}} {2}\big)
 =\sum_{r=1}^{\infty} \frac{(-1)^{r+1}} {r}{2r-1 \choose r-1} z^r
\end{equation}
and for the virial expansion
\begin{equation}
 p(\rho)={\rm ln}(1-\rho)-{\rm ln}(1-2\rho) =\sum_{r=1}^{\infty} \frac {2^{r}-1} {r} \rho^r 
\end{equation}
 
\section{Series analyses}
\subsection{ The  nonphysical singularity in the  complex activity  plane} 
Let us now observe that the LA expansions of the pressure are well
suited to estimate numerically the nearest singularity $z_d^u$ in the
plane of the complex activity $z$, which lies on the real negative
axis and thus is {\it nonphysical}, but controls the asymptotic
behavior of the coefficients and therefore the convergence radius of
the LA expansion. In $d$ dimensions the exponent $\phi_d$ of this
singularity is related\cite{fishuni,laifish} to the exponent
$\sigma_d=\phi_d -1 $ of the universal Yang-Lee edge-singularity of a
ferromagnetic spin system on a lattice of the same dimension
$d$. Using our LA expansions, we can compute the locations and
exponents of $z_d^u$ with fair accuracy for all values of $d$, either
with the aid of series-extrapolation prescriptions employing ratios of
coefficients\cite{zinn,guttda} or by Pad\'e approximants (PA) or by
differential approximants\cite{guttda}(DA). In the calculation of the
exponent, it is convenient first to locate the singularity as
accurately as possible and then use the data so obtained to form
biased PAs or DAs.  Our estimates for the location $z_d^u$ of the
nonphysical singularity and its exponent $\phi_d$, generally obtained
by first-order DAs using 15th order series on the $hsc$ lattices with
$d < 5$ or  13th order series otherwise, are shown in the Table
\ref{tab1} for a few values of $d$.  For the one-dimensional lattice
one has exactly $z_1^u=-1/4$ and $\phi_1=1/2$.  In the case of the
$sq$ lattice, the high-precision estimates $z_2^u= -0.119338886(5)$
and $\phi_2=0.83333(2)$ have been
obtained\cite{chan,jensen,gutt87,todo} by DAs, using the known 92
coefficients of the LA expansion. For $d>2$, the results cannot reach
a comparable accuracy. Completely consistent estimates of $z_d^u$ are
always obtained also from the analysis of the LA expansions of the density
or of the compressibility.

\begin{table}[H]
\scriptsize

\caption{ Estimates of the location of the nearest nonphysical
  singularities $z_d^u$ of the pressure in the complex activity plane 
  and the corresponding estimates of their exponents $\phi_d$ for
  (hyper-)simple-cubic lattices of various dimensions. 
  For $d=2,3,4$,  (even when longer expansions are known), we have reported our 
  estimates  obtained using only the first 15
  coefficients, while  for $d>4$ we have used our 13th order expansions.
  For comparison, in the second row we have reported  early
  estimates of $z_d^u$ obtained for $d=2$ and $d=3$ in
  Ref.[\onlinecite{fisgaunt,gaunt}] from series extending to the 13th
  and the 11th order respectively.  The third and fourth rows list the
  most accurate estimates of $z_d^u$ and $\phi_d$  for $d=2$ from Refs.
  [\onlinecite{chan,jensen,gutt87,hering}]. The fifth row reports our
  estimates of $\phi_d$ from the LA series, the sixth row the
  estimates of $\phi_d$ obtained from our high-order LA 
  expansions\cite{bperyl} of the hard-dimer density.}

\begin{tabular}{|c|c|c|c|c|c|c|c|c|c|c|}
 \hline
  & $d=1$  & $d=2$& $d=3$& $d=4$ &$d=5$&$d=6$&$d=7$&$d=8$\\
 \hline
 $z_d^u$ This work 
&-0.25 &-0.11934(1)& -0.07448(4)&-0.05329(5)&-0.04126(8)&-0.03365(8)
&-0.0284(1)&-0.0247(1)\\
  $z_d^u$ Ref.[\onlinecite{fisgaunt,gaunt}]&  &-0.1194(2) &-0.0744(1)&&&&&\\
  $z_d^u$ Refs.[\onlinecite{chan,jensen,gutt87,hering}]&&-0.119338886(5)&
&&&&&\\
 \hline
  $\phi_d$ Refs.[\onlinecite{chan,jensen,gutt87,hering}]&&0.83333(2)    &
&&&&&\\
 $\phi_d$ This work&0.5 &0.8335(3) &1.08(2)&1.26(5) &1.40(5) &1.45(5)&
1.50(5) &1.50(5)\\
 $\sigma_d+1$ Ref.[\onlinecite{bperyl}]&0.5&0.8338(5)&1.077(2)&1.258(5)
&1.401(9)&1.46(5)&1.495(8)&\\
 \hline
 \end{tabular} 
 \label{tab1}
\end{table}

In the case of the $hc$ lattice, we have estimated $z_{hc}^u= -0.154717(2)$
using the 25th order LA expansion presently available. The
corresponding Yang-Lee exponent is $\phi_{hc} = 0.833(2)$.

For the $hpo$ lattice, our estimate is $z_{hpo}^u=-0.1498(1)$, with
exponent $\phi_{hpo}=1.08(2)$, based on the known 23rd order LA
expansion. From the $diamond$ lattice pressure series (17
coefficients), we get $z_{dia}^u=-0.1094(2)$, with exponent
$\phi_{dia}=1.08(2)$.

In the case of the $bcc$ lattice, we estimate $z_{bcc}^u=-0.05662(2)$
using the 13th order LA expansion presently available. The exponent
estimate is $\phi_{bcc}=1.08(2)$.  Our estimate of $z_{bcc}^u$ has to
be compared with the earlier estimate\cite{gaunt}
$z_{bcc}^u=-0.0565(1)$, based on a series shorter by two terms.

All the above estimates are in fair agreement with the current most
accurate values  of the Yang-Lee edge-exponents
obtained\cite{bperyl} from 24th order hard-dimer density LA expansions,
over the same range of values of $d$, but they are subject to larger
uncertainties as they result from an analysis of shorter series.

\begin{figure}[tbp]
    \begin{center}
        \leavevmode
        \includegraphics[width=4.2 in]{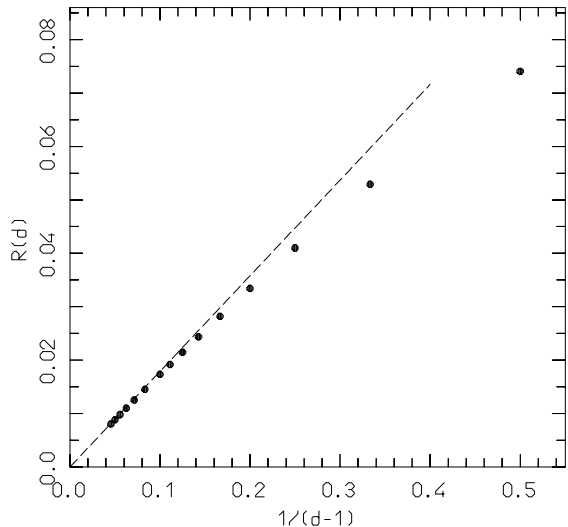}
        \caption{ \label{figura1sud} The estimated radius of the
          convergence disk of the pressure LA expansion $R(d)=-z_d^u$
          for $d=2,3,...,23$ is plotted vs $1/(d-1)$ and compared with
          its expected asymptotic behavior for large $d$ (dashed
          line).  }

    \end{center}
\end{figure}

\subsection{The large $d$ behavior of $z^u_d$} 
The structure of the $d$-dependence of the LA and the LD coefficients
in the expansions on the $hsc$ lattices is very suggestive also as far
as the large $d$ behaviors are concerned. Observing that on the $hsc$
lattices the LA expansion coefficients $c_n(d)$ are polynomials in $d$
of order $n-1$ and that the convergence radius $|z_d^u|$ of the LA
expansion equals the large $n$ limit (assuming that it exists) of the
absolute value of the ratios $|c_n(d)/c_{n+1}(d)|$, we can argue that
for large $d$, one has $z_d^u \approx a/d$ with $a=-0.179(5)$. In a
slightly better approximation, one can write $z_d^u \approx a/(d-1)$.
In the Fig. \ref{figura1sud} we have plotted our estimates of $|z_d^u|$
vs $1/(d-1)$ and compared them with their expected asymptotic behavior
for large $d$ (dashed line).

\subsection{The physical singularity in the activity}
\normalsize It is not as straightforward to estimate from our LA
expansions the critical values of $z_d^c$ associated to the physical
phase-transition.  Unfortunately, for $d>2$ our series are not sufficiently
long to locate accurately the singularities $z_d^c$ and characterize
them as $d$ varies, because $|z_d^c/z_d^u| \gg 1$, as shown by the best
current estimates\cite{hering}  obtained by a
cluster MC simulation with reduced critical slowdown, and reported in
Tab. \ref{tab2} for a few values of $d$. Very accurate estimates
of the critical exponents were also achieved in the same study, 
confirming the Ising-like nature of the transition.

In the case of the one-dimensional lattice, the pressure is analytic
for finite positive $z$ (equivalently for $\rho<1$).  For the $sq$
lattice, an extremely long LA expansion of $p(z)$ is known, and thus a
very accurate estimate of $z_2^c$ could be obtained\cite{chan}, in
spite of the fact that the physical singularity lies well outside the
convergence disk of the expansion.

  HA expansions in the
variable $1/z$ combined with LA expansions, have also proved
helpful\cite{fisgaunt,gaunt} in the study of the $d=2$ and $d=3$
cases. However HA expansions sufficiently long for this kind of studies
 are not yet known for $d>3$.

 The existence of a phase transition in the HSLG was first proved in
 Ref.[\onlinecite{dobru}]. More recently\cite{galvin}, the bound
 $z_d^c = O(d^{-1/4} {\rm ln}^{3/4} d)$ has been proved for the $hsc$
 lattices in $d$ dimensions.  Probably this bound is not optimal.  The
 empirical asymptotic formula $z_d^c \approx
 \frac{e}{2(d-1)}+O(\frac{1}{d^2})$ for large $d$, agreeing
 qualitatively with the mean-field approximation, has been
 devised\cite{hering} to fit the estimates in Tab. \ref{tab2}.  In the
 mean field approximation also $\rho_d^c$ has a similar asymptotic behavior.
 
 Studying term by term the large $d$ behavior of the virial expansion,
 we can moreover argue that the equation of state should take the form
 $p=F(\bar \rho^2)$, with $\bar \rho= \rho \sqrt d$ and that the first
 few terms of the expansion of $F(\bar \rho^2)$ in powers of $\bar
 \rho^2$ should be $F(\bar \rho^2)=\bar \rho^2-\frac{3}{2}\bar \rho^4
 -\frac{10}{3}\bar \rho^6 -\frac{147} {3}\bar \rho^8 -\frac{4536} {5}
 \bar \rho^{10} -\frac{64152} {3} \bar\rho^{12} +...$.  Consistently
 with the behavior of $\rho_d^c$, that vanishes in the large $d$
 limit, the remark suggests that the equation of state is trivial in
 this limit.

\subsection{The critical values of  the density 
 from the virial expansion} 
In the case of the one-dimensional lattice, 
  the nearest singularity
 of the pressure in the complex-density plane occurs at $\rho=1/2$, 
just on the upper border of
 the physical range $0 \le \rho \le 1/2$. 
In dimension $d=2$,  the convergence radius of the
 LD expansion of the pressure is determined by a complex-conjugate pair of
 nonphysical singularities in the right-hand half-plane of the complex
 density, whereas the physical critical value\cite{hering} of the density
 lies somewhat farther. 

 This situation occurs in the {\it hard-square}, the {\it hard-triangle} 
and the {\it hard-hexagon} model.  Unlike the $hc$
  and $sq$ lattice cases, the latter model has been solved
  exactly\cite{baxtri,baxbook} and therefore also the nonphysical
  singularities can be  located very accurately.

\begin{figure}[H]
    \begin{center}
      \leavevmode
      \includegraphics[width=4.2 in]{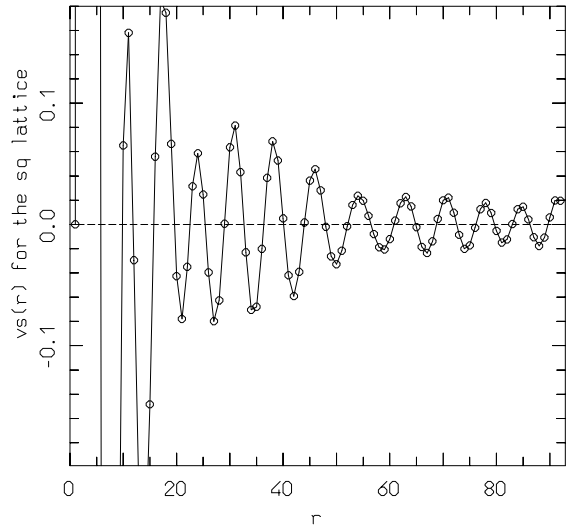}
        \caption{ \label{figurasq} The $r$th coefficient $vs(r)$ of the
          virial expansion (in terms of a density variable normalized
          with the modulus $|\rho^{nst}_2| = 0.316$ of the nearest
          singularity in the complex density plane) vs the order $r$
          of the coefficient, in the case of the $sq$ lattice. The
          values of the successive coefficients (open circles) have
          been interpolated by straight lines to profile more clearly their
          oscillating behavior. The overall normalization of the curve
          is arbitrarily chosen for graphical convenience. }
    \end{center}
\end{figure}

  For the $sq$ lattice, we have reanalysed the first 48 coefficients
  of the LD series looking for the singularities of smallest modulus
  in the PAs and DAs and have thus obtained the estimate
  $\rho^{nst}_2=0.201(5) \pm i0.244(5)$, with modulus $|\rho^{nst}_2|
  = 0.316(6)$. This result is completely compatible with the early
  estimate\cite{fisgaunt} $\rho^{nst}_2 \approx 0.21 \pm i0.26$ from a
  12th order series. On the other hand, an accurate recent MC
  estimate\cite{hering} of the critical density is $\rho^c_2 =
  0.367743000(5) > |\rho^{nst}_2|$.  It is known\cite{guttda,mcoy}
  that, if the density variable is normalized to the convergence
  radius $|\rho^{nst}_2|$ of the pressure expansion, the LD expansion
  coefficients will exhibit asymptotically regular oscillations of
  constant amplitude and period $L = 2\pi/|\theta|$ inversely
  proportional to the modulus of the phase $\theta$ of the nearest
  singularity. Using the first 92 coefficients of the long LD
  expansions reported in Refs.[\onlinecite{baxter,chan}], we have
  plotted in Fig. \ref{figurasq} the $r$th series coefficient in terms
  of the variable $\rho/\rho^{nst}_2$ vs its order $r$ to exhibit the
  first few oscillations. Consistently with our estimate of
  $\rho^{nst}_2$, an oscillation period tending to $L \approx 7$ is
  clearly discerned in the plot.

\begin{figure}[H]
    \begin{center}
        \leavevmode
    \includegraphics[width=4.2 in]{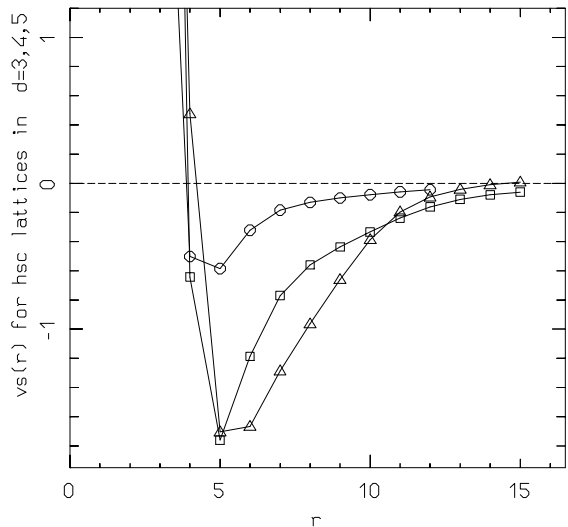}
        \caption{ \label{figura2} Same as Fig. \ref{figurasq}, for the
           coefficients of the virial expansions in the cases of the
           $hsc$ lattices in $d=3$ (open triangles), $d=4$ (open
           squares), and $d=5$ (open circles).  For the various
           curves, the density variables are normalized respectively
           with $|\rho^{nst}_3| = 0.21 $, $|\rho^{nst}_4| = 0.143$ and
           $|\rho^{nst}_5|= 0.11 $, using the estimates of
           Ref. [\onlinecite{hering}], which are more accurate than ours. 
       The overall normalization of
           each curve is arbitrarily chosen for graphical
           convenience. }
    \end{center}
\end{figure}

Similarly for the $sc$ lattice, we have plotted in Fig. \ref{figura2}
the 15 known LD coefficients vs their order. The plot might either
look barely sufficient to guess the onset of a first oscillation, with
a period longer than in the $d=2$ case, or on the contrary to exclude
any oscillation.  The PA and DA study of the series is similarly
inconclusive, suggesting that the nearest singularities are a complex
conjugate pair at $\rho^{nst}_3 \approx 0.20(6) \pm i 0.01(8)$, with
modulus $|\rho^{nst}_3| \approx 0.20$.  Thus, if oscillations occur in
the coefficient plot, their period has to be large, so that they cannot
be visible in a plot showing only the first 15 coefficients. 

In the case of dimension $d=4$, using 15 coefficients, we estimate
$\rho^{nst}_4=0.12(4) \pm i0.01(8)$ and again see no apparent oscillation of
the coefficient plot. In the case of dimension $d=5$, in which 13
coefficients are known, we obtain $\rho^{nst}_5=0.12(4) \pm i0.02(8)$
and for $d=6$, we have $\rho^{nst}_6=0.11(4) \pm i0.001(1)$. For the
various dimensions, compatible estimates are also obtained by
considering  the values of $\rho$ at which
the curvature of the pressure vanishes. The corresponding values of
$z_d$ are determined from the $\rho=\rho(z)$ curves. No unbiased
exponent estimate is possible using our expansions.

Here  it is also  worth
remarking that the virial coefficients $v_r(d)$, whose expressions for
the $hsc$ lattices are the polynomials in $d$ reported in Appendix A1, are
negative for all $d \ge 4$, and $4 \le r \le 13$.

\begin{figure}[H]
    \begin{center}
        \leavevmode
      \includegraphics[width=4.2 in]{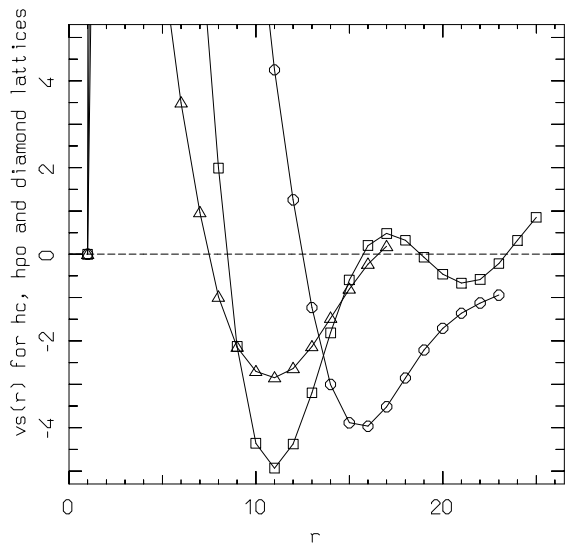}
        \caption{ \label{figura3} Same as Fig. \ref{figurasq}, for the
           coefficients of the virial expansions in the cases of the
           $hc$ lattice (open squares), the $hpo$ lattice (open
           circles) and the $diamond$ lattice (open triangles). The
           density variables are normalized respectively with
           $|\rho^{nst}_{hc}| = 0.38$,  $|\rho^{nst}_{hpo}| =0.36$
           and $|\rho^{nst}_{diam}| =0.31$.
           The overall normalization of each curve is arbitrarily
           chosen for graphical convenience. }

    \end{center}
\end{figure}

 \begin{table}[H]\scriptsize
   \caption{ In the first line, we have reported estimates of the
     critical activities $z_d^c$ for various $d$-dimensional $hsc$
     lattices obtained in Ref.[\onlinecite{bloetez,hering,jensen}] by a
     cluster MC simulation.  The second line contains the estimates of the
     critical densities $\rho_d^c$, obtained by cluster MC methods in
     Ref.[\onlinecite{bloetez,hering}] for various (hyper)-simple-cubic
     lattices of dimension $d$. In the third line, we list our estimates of
     $z_d^{nst}$ derived from the locations of the nearest singularity
     $\rho_d^{nst}$ in the density plane. The critical densities are determined
     by second-order DAs using expansions of order 48, 15, 15, 13, 13 for
     $d=2,3,4,5,6$ respectively and  are reported in the fourth line. 
     For $d>2$,  our
     estimates of the nearest singularity  seem to be 
      consistent with those of the critical densities
     reported in the second line. }
 \begin{tabular}{|c|c|c|c|c|c|c|c|c|}
  \hline
 &$d=1$  & $d=2$& $d=3$& $d=4$ &$d=5$&$d=6$\\
  \hline
   $z_d^c$ Ref.[\onlinecite{hering,bloetez,jensen}]
 & & 3.79625517391234(4)
 &1.05601(3)&0.58372(1)& 0.40259(1)&0.308217(6) \\ 
  $\rho_d^c$ Ref.[\onlinecite{hering}]&  &0.367743000(5) &0.210490(3)
  &0.143334(3)&0.109392(2)&0.088948(2) \\ 
   $z_d^c$ This work
 & $\infty$ & 
 &1.1(3) &0.7(2) &0.5(2) &0.3(1) \\ 
   $\rho_d^{nst}$  This work& $0.5$  & $0.196(15) \pm i0.243(15)$ 
 &$0.21(3) \pm i 0.01(8) $& $0.15(3) \pm i0.01(10)$&$0.11(3) \pm i0.02(8)$
 & $0.09(3) \pm i0.000(1)$\\
  \hline
  \end{tabular} 
  \label{tab2}
 \end{table}
\normalsize

From the results summarized above, we are led to conjecture that, in
the case of the $hsc$ lattices, either

{ i) only for $d =2 $, the complex pair of singularities in the
  density plane coexists with the {\it farther} singularity on the
  real axis associated to the physical phase transition, while for $d
  > 2$ these singularities approach the real axis until they pinch
  it. For larger values of $d$, either they coalesce with the physical
  singularity or replace it.  Should this picture be true, the
  physical singularity on the positive density axis would become the
  nearest one for $d \ge 3$ and thus even LD expansions of a moderate
  length might be of some value in locating it; }

or

{ ii) just like in the $d=2$ case (and perhaps also for $d=3$), it is  a
  complex-conjugate pair of singularities in the density plane
 that  determines the convergence radius of the LD expansion also for 
  values of $d \ge 3$. They will approach the real axis asymptotically as
  $d$ grows and thus the coefficients will show oscillations of
  increasingly longer, but finite period. Therefore series
  significantly more extensive than those presented here, would be necessary
  both to exhibit these oscillations and to locate reliably the
  critical value of the density $\rho_d^{nst}$.}

\normalsize To discriminate between the two possibilities, we have
compared in Tab.\ref{tab2} our series estimates for the nearest
singularities in the complex density plane $\rho_d^{nst}$, from first-
and second-order DAs, with the estimates of the critical density
$\rho_d^c$, obtained in Ref.[\onlinecite{hering}]. Of course, for
$d>2$ the accuracy of the computations using our moderately long
series is still poor, and unfortunately no simulation study has been
attempted to follow the movement of the possible nearest complex
singularities for $d>2$.  For $d=2$, it is well established that
$|\rho_2^{nst}| < \rho_2^c $. For $d \ge 3$, we observe that our
numerical estimates suggest that $|\rho_d^{nst}| \approx \rho_d^c$,
admittedly within large uncertainties.  This is consistent with the
features of the coefficient plots for the $hsc$ lattices in dimensions
$d= 3, 4, 5$ shown in Fig. \ref{figura2}.  Thus the results of the
analysis of our LD series, although not yet allowing sharp
conclusions, seem to support the first of the above pictures and
encourage us to believe that a reasonable extension of the LD series,
might be numerically useful.

We have similarly studied also the LD expansion for the $hc$ lattice,
the other 2$d$ model on a bipartite lattice for which a LA expansion
is available, and determined the nearest singularity pair:
$\rho^{nst}_{hc} \approx 0.33(2) \pm i 0.20(3)$, with modulus
$|\rho^{nst}_{hc}| \approx 0.38$. The estimates of the physical
critical parameters $z^c_{hc}=7.92(8)$ and
$\rho^c_{hc}=0.422(10)>\rho^{nst}_{hc}$ of an early
study\cite{runnsalv}, are consistent with this result. Thus, as shown
in Fig. \ref{figura3}, we expect to see an oscillation of period
tending to $L \approx 11$ in the plot of the expansion coefficients
(expressed in terms of a density variable normalized to the
convergence radius).

On the other hand, in the case of the $3d$ $hpo$, the study of the LD
series and the behavior of the LD coefficients, shown in
Fig. \ref{figura3}, suggest the location of the nearest singularity in
the complex density plane $\rho^{nst}_{hpo} \approx 0.36(1) \pm i
0.001(1)$. Consistently the PAs on the compressibility yield
$z^c_{hpo}=5.5(5)$ and $\rho^{c}_{hpo}=0.36(1)$. Similarly for the
$diamond$ lattice, we have $\rho^{nst}_{dia}= 0.31(2) \pm i 0.001(1)$
and consistently from PAs on the compressibility we have
$z^c_{diam}=2.8(2)$ and $\rho^{c}_{dia}= 0.31(2)$.  In both cases the
physical singularity in the density plane seems to be the nearest one
and therefore it can be determined, though still with a significant
uncertainty, by the LA and LD expansions. It would be interesting to
confirm these estimates by simulations.

In the case of the $bcc$ lattice, accurate MC
estimates\cite{bloetebcc} of the critical activity and of the
corresponding critical density are available: the values $z^c_{bcc}=
0.72020(4)$ and $\rho^c_{bcc}=0.1714(1)$ have been indicated by an MC
cluster algorithm. The 13 available LD series-expansion coefficients
are too few to give an accurate estimate of the nearest singularities,
but, as in the cases of the $sc$, the $hpo$ and the $dia$ lattices,
 do not suggest that they are complex.

\subsection{The hard-sphere entropy constants}

We can also show that for not too large $d$, reasonably long LA
expansions of the pressure can help
to estimate heuristically  the constants $h_d$ that control the
exponential growth of the number of all possible hard-sphere
arrangements over the $N$ sites of a finite $d$-dimensional $hsc$ lattice in
the large $N$ limit. These quantities are defined by
\begin{equation}
 h_d= S(z)|_{z=1}= \lim_{N \rightarrow \infty} \frac{1} {N}{\rm
 ln}(\Xi_N(z))|_{z=1}=p(1)
\label{entropy}
\end{equation}
and are often called {\it specific-entropy constants of the 
hard-sphere gas} for the
lattice under study.  It is easy to show that $p(1)$ is
 the maximum value of the entropy defined by Eq. (\ref{entro}). 

 The constants $h_d$ are of interest also in other fields.  In
 information theory, the quantity $h_d$ is called {\it
   capacity}\cite{nagy} of certain $d$-dimensional constrained codes
 used in digital recording applications.  As briefly anticipated in
 Sect. II, the quantity $\Xi_N(1)$, which can be defined for a general
 graph with $N$ vertices and counts the total number of distinct
 ``independent sets of vertices'' i.e. of subsets of pairwise
 non-adjacent vertices, usually called
 ``vertex-independence-number''\cite{calkin} (or Fibonacci
 number\cite{engel} of the graph) is of interest both in combinatorial
 mathematics\cite{finch} and in theoretical chemistry. In the latter
 context it is called Merrifield-Simmons index\cite{merri} and used
 for a topological characterization of a (large) molecule whose
 structure is represented by the graph.

We have recently studied\cite{buperg} a Grassmann-algebra algorithm
that can efficiently compute the ``independence polynomial'' even for
relatively large graphs and in particular for finite lattice
graphs. Moreover, extrapolating to infinite lattice the results of
this algorithm by the prescriptions of the
``finite-lattice-method''\cite{flm}, we can reproduce at least part of
the very long already known LA expansion\cite{baxter,chan} for the $sq$
lattice.  Unfortunately our algorithm cannot yet compete with the transfer
matrix and thus we cannot extend the existing results.

\begin{figure}[H]
    \begin{center}
        \leavevmode
       \includegraphics[width=4.2 in]{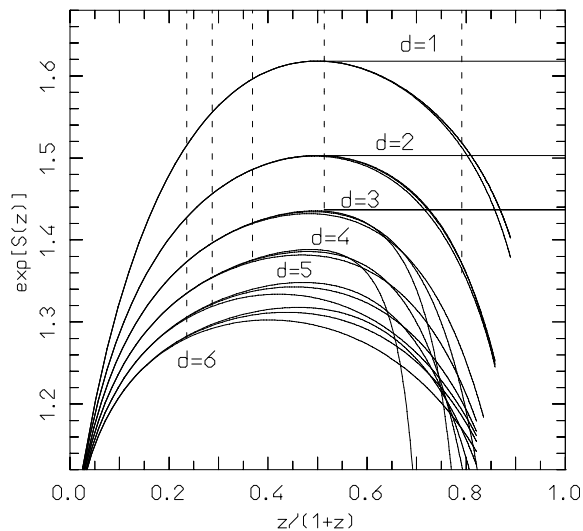}
        \caption{ \label{fig_entr1} The exponential of entropy, for the
          $hsc$ lattices, is plotted vs $z/(1+z)$ for $d=1,2,3,...,6$.
          This quantity is computed by quasi-diagonal PAs formed with
          15 series coefficients in dimension $d=1,2,3,4$ and with 13
          coefficients in $d=5,6$.  The
          vertical dashed lines indicate the values of
          $z_d^c/(1+z_d^c)$, for each $d$. For $d=1$ the exact result,
          and for $d=2,3$ good approximations for the expected values
          of $exp[S(z)]$ at $z=1$ are indicated by horizontal
          continuous lines.}
    \end{center}
\end{figure}

No closed form expressions are known for the quantities $h_d$, except
for $d=1$ i.e. $h_1= {\rm ln}((1+\sqrt 5)/2)$, but for $h_2$, $h_3 $ and
$h_4$ rigorous lower and upper bounds have been obtained.  In the
thoroughly studied\cite{bax,chan,todo,jensen} case of the $sq$
lattice, a high-precision determination\cite{bax} of $\exp(h_2)
=1.5030480824753322643220663294755536893857810...$ was achieved by
extrapolating to infinite lattices variational results from the
corner-transfer-matrix method. This value is probably
correct\cite{bax} to 43 decimal places, and is clearly consistent with
the rigorous bounds\cite{calkin,nagy,golin,gamarnik}
 indicated in Table \ref{tab4}. The value of the density at $z=1$ is
 $\rho_2(1)=0.22657081546271468894199226347129902640080...$, probably
 correct\cite{bax} through the 41 decimal figures reported here.  Further 
improvement of the above estimates by the presently available LA
 series seems to be out of question. Here we shall be interested only
 into the higher values of $d$ and use some of the long $d=2$
 expansions mainly to understand the limitations of a simple
 series resummation method and to support our claim that, at least for not
 too large $d$, reasonable estimates of $h_d$ can be obtained even
 from the moderately long series derived in this paper. To begin with,
 we show how accurate are the estimates that we obtain for $d=2$. For
 example, on the $sq$ lattice a naive [7/7] PA which uses only the
 first 15 coefficients of the LA expansion of $p(z)$, (or better
 first-order DAs using the same set of coefficients), can reproduce
 correctly the first four digits of $\exp(h_2) = 1.50305...$.  Using
 an increasing number of the many coefficients known for the $sq$
 lattice, we observe that $p(1)$ and $\rho(1)$, determined in this
 way, reproduce an increasing number of figures of the above mentioned
 values and can argue that roughly one more digit of the expected
 values can be gained for every two additional coefficients used in
 forming the approximants.  Thus it is tempting to test the accuracy
 of this simple series approach also with the cubic lattices in $3d$
 and $4d$, in which only the first 15 coefficients of the LA expansion
 are known and even to extend this test to higher values of $d$.  From
 the highest PAs and DAs, we get the estimate $exp(h_3) = 1.4366(3)$,
 which is marginally compatible with the known\cite{fried2} bounds:
 $1.4365871627266 \le \exp(h_3) \le 1.43781634614$.  For the
 hyper-simple-cubic lattice in $4d$ (again using 15 coefficients), our
 DA estimate is $exp(h_4)=1.394(-5)(20)$. The central value of this
 estimate lies approximately $2\%$ below the lower end of the
 range\cite{metcalf,gamarnik} defined by the rigorous inequalities
 $1.417583 \le \exp(h_4) \le \exp(h_3) \le$ upper bound of $
 \exp(h_3)= 1.4378..  $.  This upper bound improves marginally the
 best previous one $1.4447..$ obtained in
 Ref.[\onlinecite{gamarnik}]. Even so, the bounds for $d= 4$ remain
 less tight than for $d<4$.  Finally, using the 13 coefficients
 derived in dimension $d=5$, DAs predict $\exp(h_5) = 1.36(-1)(+4)$.
 Unfortunately, in this case no bound tighter than $exp(h_5)<exp(h_4)$
 is presently available\cite{metcalf}, but we can reasonably suppose
 that our DA results underestimate the correct value by less than
 $5\%$.  These results indicate that the simplest PAs or DAs can yield
 relatively accurate analytic continuations of the $z$-expansion of
 $p(z)$ well outside their convergence disks (of radius $|z_d^{u}|$),
 but also that the precision of this procedure deteriorates rapidly
 with increasing space dimension $d$. One reason for this is the
 shrinking of the convergence radius of the LA expansion (see Table 
 \ref{tab1}). Another reason is the decrease
 of $z_d^c$ as $d$ grows (see Table \ref{tab2}). One should also
 consider that, for $d>4$, the LA series extend only to 13th
 order. Since the pressure is expected to be continuous, although
 non-analytic at $z_d^c$, one can however hope that longer series
 might make more accurate estimates possible. Notice for example that,
 already for $d=3$ we have $z^c_3 \approx 1.056$, while $z_d^c < 1$
 for $d \ge 4$.  To show how this loss of accuracy as $d$ grows is
 related to the location of $z_d^c$, we have plotted in
 Fig. \ref{fig_entr1} the quantities $exp[S(z)]$ vs $z/(1+z)$ for $hsc$
 lattices of dimension $d=1,...,6$.  They are computed by the PAs of
 orders [7/7], [6/6] and [5/5] for $d=1,2,3,4$ and of order [6/6] and
 [5/5] and [4/4] for $d=5,6$, formed with the LA expansions of lengths
 15 in the first case and 13 in the second. We have marked by vertical
 dashed lines the values of $z_d^c/(1+z_d^c)$ to show that, when $d
 \ge 3$, the convergence of the PAs, that can be judged from the
 spread of their values for $z \gtrsim z_d^c$, begins to deteriorate
 before the maxima of the curves are attained, thus making the
 approximations less reliable.  For $d=2$ and $d=3$, the known upper
 and lower bounds of $exp\big [S(z)|_{z=1} \big ]$ suggest good
 approximations of this quantity, indicated in the figure by
 horizontal continuous lines. For $d=4$, the available bounds are too
 loose to be useful in the assessment of the accuracy and we have not
 indicated them in the figure.

 Our estimates of $\exp(h_d)$ for cubical lattices of dimensions
 $d=2,3,...,5$ are summarized in Table \ref{tab4} and compared with
 the upper and lower bounds, whenever available. In the same table we
 have also reported our estimates of the hard-sphere densities
 $\rho_d(1)$ in the same approximation.  Our central estimates quoted
 in the tables are obtained as the averages of the highest-order
 near-diagonal PAs (or DAs) of $p(z)|_{z=1}$, that can be formed from
 the available LA expansions. To the estimates we have attached
 uncertainties not smaller than twice the maximum variation among the
 PA values.

\subsection{Entropy constants for other lattices}

 Transfer-matrix computations\cite{bax} have yielded 
 high-accuracy estimates also in the case of the $hc$ lattice. For
 the entropy, they give $\exp(h_{hc}) \approx
 1.54644070878756141848902270530472278026... $, and for the density in
 $z=1$, they give $\rho_{hc}(1) \approx
 0.2424079763616482188205896378263422541...$. Both values are probably
 correct through the figures reported.  Similarly to our previous
 results, in the case of the $hc$ lattice, a [7/7] PA estimate
 $\exp(h_{hc}) = 1.54642(2)$, reproduces correctly the first four
 digits of the above value, while the [12/12] PA (that uses all the 25
 known coefficients) yields $\exp(h_{hc}) = 1.54644069(2)$ reproducing
 the first 8 digits. A similar precision is achieved for the
 corresponding density: the [7/7] PA gives $\rho_{hc}(1)= 0.24240(1)$
 and the [12/12] PA gives $ \rho_{hc}(1)=0.2424079(3)$.
 Moreover, these estimates of $\exp(h_{hc})$ are consistent with the
 loose inequalities\cite{metcalf} $exp(h_1)>exp(h_{hc})>exp(h_2)$,
 where $\exp( h_1) = (1+\sqrt 5)/2 \approx 1.6180...$.  

No earlier estimates are available   for the $hpo$ lattice. We obtain
 $exp(h_{hpo}) =1.54564(3) $ using a [7/7] PA, while an [11/11] PA yields
 $exp(h_{hpo}) = 1.545659(1) $. The corresponding approximations for
 the density are $\rho_{hpo}(1)=0.24114(2)$ and
 $\rho_{hpo}(1)=0.241153(1)$, respectively. Following 
 Ref.[\onlinecite{metcalf}], we can expect that $exp(h_{hpo})>exp(h_3)$.

\begin{figure}[H]
    \begin{center}
        \leavevmode
       \includegraphics[width=4.2 in]{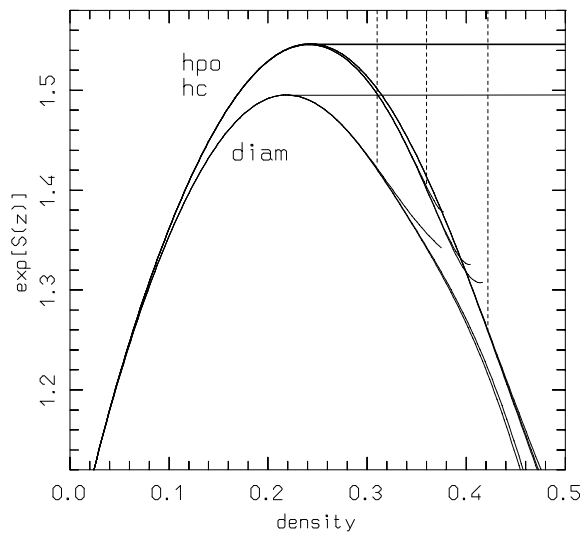}
        \caption{ \label{fig_entr2} The exponential of entropy,
          computed by the highest order available quasi-diagonal PAs,
          is plotted vs the density $\rho$ for the $hc$, the $hpo$ and
          the $diamond$ lattices. We have indicated by vertical dashed
          lines the values of $\rho_c$ corresponding to the various
          lattices. Notice that the curves for the $hc$ and the $hpo$
          lattices are indistinguishable on the scale of the figure
          except in the right hand region.}
    \end{center}
\end{figure}

 In the case of the $diamond$ lattice, we estimate
 $exp(h_{dia})=1.49526(6) $ and $\rho_{dia}(1)=0.2188(3)$ and can
 expect\cite{metcalf} that $exp(h_{dia})>exp(h_3)$.  We are not aware
 of tighter bounds for these lattices.  Notice that for the $hc$, the
 $hpo$ and the $diamond$ lattices, the critical values of $z$ are $ \gg 1$
 and thus the accuracy of our entropy estimates is justified. Our PA
 results for the $hc$, $hpo$ and $diamond$ lattices are shown in
 Fig. \ref{fig_entr2}, which differs from Fig. \ref{fig_entr1} only
 because $exp[S]$ is plotted vs $\rho$ instead of $z/(1+z)$.

In the case of the $bcc$ lattice, we
 estimate $exp(h_{bcc}) = 1.41(2)$ and correspondingly
 $\rho_{bcc}(1)=0.21(1)$.  The expected inequality\cite{metcalf}
 $exp(h_{bcc}) < $ upper bound of $exp(h_3)$ is satisfied.  
\begin{table}[H]\scriptsize
\caption{ DA estimates of the hard-sphere entropy constants
 $h_d=p(1)$ and the corresponding estimates of the densities
 $\rho_d(1)$ for (hyper)-simple-cubic 
lattices of various dimensions. In the second column we have
reported tight lower bounds and in the fourth tight upper bounds for $h_d$
on the $hsc$ lattice, when available\cite{nagy,calkin,golin,fried2,gamarnik}.  
}
\begin{center}
\begin{tabular}{|c|c|c|c|c|c|}
 \hline
 $hsc$ & Lower bound& This work& Upper bound   \\
 \hline
  $\exp (h_2)$  &1.503047782 &1.50305(2)&1.503058  \\
  $\exp (h_3)$  &1.4365871627266 &1.4366(3)&1.43781634614 \\
  $\exp (h_4)$  &1.417583 &1.394(-5)(20)&1.43781634614  \\ 
  $\exp (h_5)$  & &1.36(-1)(4)&  \\
 $\rho_2(1)$ & &0.2265(2)& \\
 $\rho_3(1)$ & &0.202(2)& \\
 $\rho_4(1)$ & &0.18(2) & \\
 $\rho_5(1)$ & &0.16(2) &\\
 \hline

\end{tabular} 
\end{center}
 \label{tab4}
\end{table}

\section{Conclusions}
Using the results of our recent study of the Ising model HT
expansions\cite{BP}, we have computed through order $13$ the LA and LD
expansions of the pressure for the HSLG model on the (hyper)-simple-cubic
lattices in $d$ dimensions, and written their expressions as
polynomials in $d$ through order $13$,
 assuming that the general virial coefficient $v_k(d)$ is
a polynomial of degree $[\frac{k}{2}]$, as confirmed by our calculations.

With the aid of the current series-extrapolation techniques, we have
then analyzed the LA and LD expansions of the pressure for the HSLG
model on the (hyper)-simple-cubic lattices, in general dimension $d$
as well as the expansions for the $hc$, the $hpo$ and the $diamond$
lattices already existing in the literature, but not thoroughly
studied.  We have thus obtained fairly accurate and nontrivial
information, in some cases not previously known, concerning the
parameters of the nonphysical singularities and their behaviors as the
lattice dimension $d$ grows.  As to the more difficult problem of
determining the physical singularities by series computations, the
presently known expansions for $d>2$ are still too short to yield
estimates that can compete with those from the simulations, but they
may help to discuss and justify our conjecture that only in the case
of two-dimensional lattices, the nearest singularity in the right-hand
plane of the density is a nonphysical complex-conjugate pair. If
valid, our conjecture might support the expectation that just somewhat
longer expansions can locate reliably the physical singularities also
for $d>2$.

We have also observed that the simplest way to estimate the entropy
constants from our series, by PA or DA resummation of the LA series,
is very effective for $d=2$ and $d=3$, but becomes less accurate for
$d>3$, probably only because the available series are not long enough.

Finally our results suggest that, if the series study of the HSLG
model for $d \ge 3$ is considered worth pursuing, a significant
extension of the LT series-expansions of the spin 1/2 Ising model in a
magnetic field, should be undertaken, because for $d>2$, the LA and LD
expansions can only be derived by following this route.

\section*{Acknowledgments} 
We thank INFN for supporting this research.

\normalsize
\section{Appendix A0: Computing the $L_n^{(1/2)}$ from 
the   HT expansions} 

  Let ${\cal
    F}(K,H) \equiv \lim_{N \to \infty} \frac {-1} {N\beta}{\rm ln} Z_N(K,H)$
  denote the specific free energy for the $S= 1/2$ Ising model
  subject to a magnetic field $H$.
  The corresponding HT series
  expansion can be written as
\begin{equation}
-\beta  {\cal F}_{HT}(K,H) = \frac{q}{2} {\rm ln}(cosh(K)) 
+ h + {\rm ln}(1 + \mu) + \sum_{i=1}^\infty \psi_i(c) t^i
\label{Bwh0}
\end{equation}
Here we have set $h=\beta H$, $c = tanh(h)$, $\mu = e^{-2h}$ and $t =
tanh(K)$.

 The specific free energy is obtained as a  HT series, observing that
 the magnetization ${\cal M}(K,h)= -\frac {\partial {\cal F}_{HT}}
{\partial H} = c + \sum_{i=1}^\infty M_i K^i$ can be computed using
the linked-cluster method\cite{wortis} in terms of the quantities
$M_i$ that are polynomials in the bare vertices $V_i$, defined by
\begin{equation}
V_{i+1}= \frac {d V_i}{d h} = ((1-c^2)\frac{d}{d c})^i c
\label{Bbarev}
\end{equation}
with $V_0= {\rm ln}2 cosh(h)$. By integrating the magnetization, one
can compute $ {\cal F}_{HT}(K,H)$ (up to an integration constant
depending on $K$) and thus can determine the quantities $\psi_i(c)$
appearing in Eq. (\ref{Bwh0}), using $\frac{d (-\beta {\cal F}_{HT})}{d c} =
 \frac{\cal M}{1-c^2}$.

The quantities $\psi_i(c)$ are  polynomials in $c$
of degree $2i$, even due to the symmetry $h \to -h$.

\noindent The $n$th order coefficient, in powers of $\mu$, of the LT expansion
for high magnetic field
results from excitations in which precisely $n$ spins are
  flipped with respect to the background with all spins up.
The LT free energy is given by Eq. (\ref{fl0})
\begin{equation}
-\beta {\cal F}_{LT} = K \frac{q}{2} + h + \sum_{k=1}^\infty L_k(u) \mu^k
\label{Bwl}
\end{equation}
where $u=\exp(-4KJ)$.
Here and in what follows, for brevity, we shall set  $L_k(u) = L_k^{(1/2)}(u)$.

In a domain around infinite magnetic field $ {\cal F}_{HT}(K,H)$ and
$ {\cal F}_{LT}(K,H)$ are both defined, so they are related by analytic
continuation\cite{ domb}; it is convenient to write the HT variables
$t$ and $c$ as
\begin{equation}
t(z) = \frac{1 - \sqrt{1-z}}{1 + \sqrt{1-z}}
\end{equation}
 with $z= 1-u$ and
\begin{equation}
c(\mu) = \frac{1 - \mu}{1 + \mu}
\end{equation}

Equating the expansions ${\cal F}_{LT} = {\cal F}_{HT}$ as $K \to 0$
 leads to the identities
\begin{equation}
L_k(1) = (-1)^{k+1}/k
\label{Bln1}
\end{equation}
while for $h \to \infty$ one has
\begin{equation}
\frac{q}{2}({\rm ln}(cosh(K)) - K) + \sum_{i=1}^\infty\psi_i(1)t^i = 0
\nonumber
\end{equation}
Therefore one gets
\begin{equation}
\sum_{k=1}^\infty(L_k(u) - L_k(1))\mu^k = 
\sum_{i=1}^\infty (\psi_i(c) - \psi_i(1))t^i
\label{Bln2}
\end{equation}

Let us denote the coefficient of the power $x^r$ in the Taylor expansion of
$f(x)$ at $x=0$ by ${\cal C}_r(f(x)) = \frac{1}{r!}\frac{d^r f}{d
x^r}|_{x=0}$, then expanding Eq. (\ref{Bln2}) about $\mu=0$  

\begin{equation}
L_k(u(z)) = L_k(1) + \sum_{i=1}^\infty {\cal C}_k(\psi_i(c(\mu)))
t(z)^i.
\label{Bln3}
\end{equation}

The Eq. (\ref{Bln3}) gives the $j$-the derivative of $L_k$ in $u=1$ in terms
of ${\cal C}_k\psi_1,...,{\cal C}_k \psi_j$. For instance
from $\psi_1 = d c^2$ one obtains the constraint
\begin{equation}
    L_k'(1) = d k (-1)^{k+1}
\end{equation}

$L_k(u)$ is a polynomial of degree $\frac{k q}{2}$ in $u$, hence the
RHS of Eq. (\ref{Bln3}) has to be a polynomial of the same degree in
$z$.  Since $t(z)=O(u)$, the infinite sum is actually  a finite
sum.  The range of the sum is further reduced observing that $L_k(u)$
is a polynomial in $u$ of degree $\frac{k q}{2}$ with the structure
\begin{equation}
L_k(u)=\sum_{r=0}^{max[N_{--}]} [k,r] u^{qk/2-r}
\label{Blk}
\end{equation}
The notation adopted for the coefficients $[k,r]$ of the polynomial
$L_k(u)$ reproduces\cite{syk25} that commonly used in the literature.
In Eq. (\ref{Blk})   $r=N_{--}$ is
the number of adjacent pairs of flipped spins in the various configurations of
$k$ flipped spin described by $L_k(u)$.

Thus $L_k(u)$ contains\cite{bell}  only the powers  $u^r$ with $r$ ranging from 
$s_k=\frac {k q}{2} -max[N_{--}]$ to $\frac {kq}{2}$.

Considering the lattice subgraph whose vertices are the flipped spins
and whose edges connect pairs of adjacent flipped spins,  it is clear
that $max[N_{--}]$ is the (lattice-dependent) maximum number of links
in this class of $k-$vertex lattice subgraphs.

Thus to compute $L_k$, it is  sufficient to know the HT expansion in a
magnetic field through the order $s_k=\frac{k q}{2} - max[N_{--}]$.

In fact, defining the truncated sum ${\cal T}_n(f(x), x) \equiv
\sum_{r=0}^n (\frac{1}{r!}\frac{d^r f}{d x^r}|_{x=0})x^r$, we get
\begin{equation}
L_k(u) = u^{s_k}{\cal T}_{kq/2-s_k}\bigg(
(1-z)^{-s_k}\Big(\frac{(-1)^{k+1}}{k} + 
\sum_{i=1}^{kq/2-s_k }
t(z)^i {\cal C}_k(\psi_i(c(\mu)))\Big), 
z \bigg)|_{z=1-u}
\label{Bln5}
\end{equation}

So far, our remarks are true independently of the lattice structure.
Let us now restrict to the $d$-dimensional (hyper)-simple-cubic
lattices for which $q=2d$.  For lattices of dimension $d=0, 1$, the
free energy and thus the $L_k$ are known at all orders.  We have been
able to compute\cite{BP} the HT magnetization exactly in terms of $c$
through the order $n_d$ in $K$, with $n_d=24$ for $2 \le d \le 4$,
$n_d=22$ for $5 \le d \le 6$, $n_d=20$ for $7 \le d \le 10$.

For $2 \le d \le 4$, we were then able to compute $L_k$ through the
following orders: 16th for $d=2$, 14th for  $d=3$ and  13th for $d=4$.

Our determination of the $L_k$ for cubic lattices of dimension $d \le 3$ agrees with
the data in the literature\cite{sykes2d3d}.  For $d=4$ we have noticed
a single misprint in Ref. [\onlinecite{sykes4d}]: the lowest degree
coefficient of $L_8$ should be $4u^{20}$ instead of $4u^{21}$. For
$d>4$, no data exist in the literature.  

In Table \ref{Bgsk},
we have listed the quantities $qk/2-max[N_{--}]$ for $d \le 4$.  

For $d \ge 4$ and $k \le 16$ it is easy to see that $k d - s_k(d)$ is
independent of $d$.  Since  a $d$-dimensional hyper-cube has
$2^d$ vertices, for $k \le 2^{d-1}$ the graphs with maximal number of
links are subgraphs of a $(d-1)$-hyper-cube, which is a face of the
$d$-hyper-cube.  Therefore $k d - s_k(d)$ is the same for the lattices
in $d$ and $d-1$ dimensions, if $k \le 2^{d-1}$.  (One can improve
this correspondence observing that the graphs with $k \le 2^{d} -
2^{d-2}$ vertices and maximal number of links lie on two adjacent
$2^{d-1}$ hyper-cubes, so that $k d - s_k(d)$ is the same for $d$ and $d-1$
if $k \le 2^{d} - 2^{d-2}$; however we will not use this remark here).
Knowing $s_k$ for $d=3$ and $k \le 8$, we can find the
corresponding values for $d=4$ and e.g. can conclude that
for $d=4$ the smallest power of $u$ in the polynomial $L_8$ is $u^{20}$.

In $d \ge 5$, for graphs with $16$ vertices or less, the vertex
configurations with maximal number of links are those on a
$4$-hypercube forming a face of the $d$-hyper-cube; therefore $k d -
s_k(d)$ is the same as in $d=4$.

In this way, we have been able to determine the polynomials $L_k$ through
 the following orders:
13th order in dimension $d=5,6$ and  12th in $d=7,8,9,10$.

\begin{table}[ht]
\caption{This Table lists the lowest powers $qk/2-max[N_{--}]$ of $u$ appearing in $L_k$
 for the $d$-dimensional (hyper)-simple-cubic lattice. The last
line gives $max[N_{--}]=d k - s_k(d)$ for $d \ge 4$, which is the HT order
whose knowledge is required to compute $L_k$ from Eq. (\ref{Bln5}).}
\begin{tabular}{|c|c|c|c|c|c|c|c|c|c|c|c|c|c|c|}
\hline
$d$/$k$& 1& 2& 3& 4& 5& 6& 7& 8& 9&10&11&12&13&14\\
\hline
1& 1& 1& 1& 1& 1& 1& 1& 1& 1& 1& 1& 1& 1& 1\\
\hline
2& 2& 3& 4& 4& 5& 5& 6& 6& 6& 7& 7& 7& 8& 8\\
\hline
3& 3& 5& 7& 8&10&11&12&12&14&15&16&16&18&19\\
\hline
4& 4& 7&10&12&15&17&19&20&23&25&27&28&30&31\\

\hline
\hline
$max[N_{--}]$& 0& 1& 2& 4& 5& 7& 9&12&13&15&17&20&22&25\\
\hline
\end{tabular}
\label{Bgsk}
\end{table}

\section{Appendix A1: The  activity and
 virial expansions for the hard-sphere gas
 on the (hyper)-simple-cubic lattice of dimension $d$}

In the case of the $hsc$ lattices, the coefficients $[k,i](d)$ of the
powers of $u$ in the LT-polynomials $ L^{(1/2)}_k(u)$ can be expressed
as polynomials in the dimension $d$, of degree $k-1$. Using their
values for $d=0,...,10$, determined as indicated in the previous
Appendix, we have been able to write them as polynomials in $d$ through $k=11$. For $k <
11$, the HT-LT relationship leaves $11 - k$ consistency checks to be satisfied among
the coefficients.

The coefficient $[k,0](d)$ of the highest-order power in $u$ of
$L_k(u)$ is the kth coefficient $c_k(d)$ in the LA expansion ({\it
  Mayer b-series}\cite{mayer}) $ p =\sum_{r=1}^{\infty} c_r(d)z^r$ of the HSLG
model pressure.

Once we have obtained the LA expansion coefficients $c_r(d)$ for $r
\le 11$, the LD expansion coefficients of the pressure ({\it Mayer
  $\beta$-series}\cite{mayer}) $p=\sum_{r=1}^{\infty} v_r(d)\rho^r$ are computed
through the same order by the standard procedure.

We can observe that for $r \le 11$, the virial coefficients $v_r(d)$
are polynomials in $d$ of degree $[\frac{r}{2}]$ and thus we are led
to assume that this property of $v_r(d)$ remains valid for any $r$.
We might argue, more convincingly that this property is a consequence
of the fact that the virial expansion uses only one-vertex-irreducible
({\it star}) Mayer graphs, whose (strong) embedding factors\cite{domb}
are themselves polynomials in $d$ of degree at most $[\frac{r}{2}]$.
A completely analogous argument\cite{isigend} is valid for the HT
expansions in general dimension $d$.  Using the coefficients
$[k,r](d)$ for $d \le 6$ and $k \le 13$, we have then extended our
computation of the activity expansion coefficients $c_k$ and hence
$v_k$ for $d \le 6$ and $k \le 13$, since according to the above
remark this is enough to determine the polynomials $v_k(d)$ for $k \le
13$.

From the virial expansion coefficients $v_k(d)$ with $k \le 13$, we obtain 
 the activity expansion coefficients $c_k(d)$:

\scriptsize
\[c_1(d)=1\]

\[c_2(d)=-d - 1/2\]

\[c_3(d)=2d^2 + d + 1/3\]

\[ c_4(d)=-16/3d^3 - 3/2d^2 - 5/3d - 1/4\]

\[ c_5(d)= 50/3d^4 - 2/3d^3 + 47/6d^2 + 7/6d + 1/5\]

\[c_6(d)=-288/5d^5 + 24d^4 - 118/3d^3 - 4d^2 + 1/10d - 1/6\]

\[c_7(d)=9604/45d^6 - 882/5d^5 + 1913/9d^4 + 23/6d^3 - 719/45d^2 + 227/30d + 1/7 \]

\[c_8(d)=-262144/315d^7 + 47104/45d^6 - 54112/45d^5 + 1079/9d^4 + 11689/90d^3
 - 4841/180d^2 - 2743/70d - 1/8 \]

\[c_9(d)=118098/35d^8 - 201204/35d^7 + 6939d^6 - 8272/5d^5 - 21497/40d^4 - 26461/60d^3\]\[
 + 174025/168d^2 - 37319/140d + 1/9\]

\[ c_{10}(d)=-8000000/567d^9 + 640000/21d^8 - 7600000/189d^7 + 139960/9d^6 
- 21262/135d^5  + 65629/9d^4 \]\[ - 12451745/1134d^3 + 294169/252d^2 + 437071/252d - 1/10\]

\[c_{11}(d) =857435524/14175d^{10} - 449976494/2835d^9 + 219497872/945d^8 - 118160977/945d^7 
+ 76322413/2700d^6\]\[  - 14910479/216d^5  + 1694726807/22680d^4 + 
910827493/22680d^3 - 96126973/1400d^2  + 22181029/1260d + 1/11 \]

\[ c_{12}(d)=-509607936/1925d^{11} + 143327232/175d^{10} - 46669824/35d^9 + 4603392/5d^8 
- 62554848/175d^7 + 13448508/25d^6\]\[  - 13292172/35d^5 - 51285963/70d^4 + 2776957507/3150d^3 
- 1684784/25d^2 - 1891097759/13860d - 1/12 \]

\[ c_{13}(d)=  551433967396/467775d^{12} - 218615833228/51975d^{11} + 323294840011/42525d^{10} 
- 18152600453/2835d^9 \]\[  + 189827943799/56700d^8 - 73090795151/18900d^7 + 1019903740273/680400d^6 
+ 111272938283/15120d^5 - 4062673959889/680400d^4\]\[  - 1208506024709/226800d^3 
+ 1408813275251/207900d^2 - 22009079311/13860d + 1/13 \]

\normalsize

For $d=3$, two additional  coefficients of the LA expansion are known\cite{sykes2d3d}

\scriptsize

$c_{14}(3)=-2544845359479$

$c_{15}(3)=445909846722299/15$

\normalsize

For  $d=4$, the additional known coefficients are\cite{sykes4d}

\scriptsize
$c_{14}(4)=-1922838922481311/14$

$c_{15}(4)= 11091476544856728/5$

\normalsize

The virial coefficients $v_r(d)$ are:

\scriptsize
\[v_{1}(d) =1\]

\[v_{2}(d) =d + 1/2\]

\[v_{3}(d) =2d + 1/3\]

\[v_{4}(d) =-3/2d^2 + 5d + 1/4\]

 \[v_{5}(d) =-8d^2 + 14d + 1/5\]

 \[v_{6}(d)=-10/3d^3 - 15d^2 + 86/3d + 1/6\]

 \[v_{7}(d) =-40d^3 + 66d^2 - 8d + 1/7\]

 \[v_{8}(d) =-49d^4 + 301/6d^3 + 469/4d^2 - 260/3d + 1/8\]

\[v_{9}(d) =-784d^4 + 11168/3d^3 - 5716d^2 + 2834d + 1/9\]

\[v_{10}(d) =-4536/5d^5 + 2406d^4 + 5379d^3 - 19314d^2 + 62692/5d + 1/10\]

\[v_{11}(d)=-53984/3d^5 + 417380/3d^4 - 1174120/3d^3 + 1409620/3d^2-199446d + 1/11\]

\[ v_{12}(d) = -21384d^6+348964/3d^5+91157/6d^4-1087185d^3+6211568/3d^2-3279380/3d+1/12\]

\[v_{13}(d)=-495936d^6+5617424d^5-24986848d^4+54097332d^3-56281176d^2+22049834d + 1/13\]

\normalsize

For $d=3$ the additional known coefficients of the LD expansion are

\scriptsize

$v_{14}(3) =-425385/2$

$v_{15}(3)=13265929/15$

\normalsize

For  $d=4$ the additional known coefficients are

\scriptsize
$v_{14}(4)= -1811619255/14$

$v_{15}(4)= -2082055915/3$

\end{document}